\documentclass[traditabstract]{aa}

\usepackage{graphicx}
\usepackage{txfonts}
\usepackage{natbib}
\bibpunct{(}{)}{;}{a}{}{,}
\usepackage{array}
\usepackage{xspace}
\usepackage{lscape}
\usepackage{url}
\usepackage{arydshln}

\newcommand{\FeKa}{Fe K\ensuremath{\alpha}\xspace}
\newcommand{\kms}{\ensuremath{\mathrm{km\ s^{-1}}}\xspace}
\newcommand{\NH}{\ensuremath{N_{\mathrm{H}}}\xspace}
\newcommand{\redchi}{\ensuremath{\chi _\nu ^2}\xspace}

\newcommand{\xabs}{{\it xabs}\xspace}
\newcommand{\xmm}{{\it XMM-Newton}\xspace}

\newcommand{\rosat}{{\it ROSAT}\xspace}

\newcommand{\eso}{{ESO~113-G010}\xspace}
\newcommand{\Ha}{H\ensuremath{\alpha}\xspace}
\newcommand{\Hb}{H\ensuremath{\beta}\xspace}

\begin{document}

\title{The X-ray warm absorber and nuclear obscuration in the Seyfert 1.8 galaxy ESO~113-G010}

\author{Missagh Mehdipour, Graziella Branduardi-Raymont, \and Mathew J. Page
}
\institute{Mullard Space Science Laboratory, University College London, Holmbury St. Mary, Dorking, Surrey, RH5 6NT, UK \\{\email{missagh.mehdipour@ucl.ac.uk}}
}

\date{Received 27 January 2012 / Accepted 12 April 2012}

\abstract{}{}{}{}{}
\abstract
{We present the first analysis of the X-ray warm absorber and nuclear obscuration in the Seyfert 1.8 galaxy {\object{ESO 113-G010}}. We used archival data from a 100 ks \xmm observation made in 2005. From high resolution spectroscopy analysis of the RGS data, we detect absorption lines originating from a warm absorber consisting of two distinct phases of ionisation, with $\log\xi\approx 3.2$ and 2.3 respectively. The higher-ionised component has a larger column density and outflow velocity ($\NH \approx 1.6 \times 10^{22}$ $\mathrm{cm}^{-2}$, $v \approx -1100\ \kms$) than the lower-ionised component ($\NH \approx 0.5 \times 10^{22}$ $\mathrm{cm}^{-2}$, $v \approx -700\ \kms$). The shape of the optical-UV continuum and the large Balmer decrement ($\rm{\Ha/\Hb} \sim 8$) indicate significant amount of reddening is taking place in our line of sight in the host galaxy of the AGN; however, the X-ray spectrum is not absorbed by cold neutral gas intrinsic to the source. We discuss different explanations for this discrepancy between the reddening and the X-ray absorption, and suggest that the most likely solution is a dusty warm absorber. We show that dust can exist in the lower-ionised phase of the warm absorber, which causes the observed reddening of the optical-UV emission, whereas the X-rays remain unabsorbed due to lack of cold neutral gas in the ionised warm absorber. Furthermore, we have investigated the uncertainties in the construction of the Spectral Energy Distribution (SED) of this object due to obscuration of the nuclear source and the effects this has on the photoionisation modelling of the warm absorber. We show how the assumed SEDs influence the thermal stability of each phase and whether or not the two absorber phases in \eso can co-exist in pressure equilibrium.}

\authorrunning{M. Mehdipour et al.}
\titlerunning{The warm absorber and nuclear obscuration in ESO 113-G010}
\maketitle

\section{Introduction}
\eso was first identified and catalogued as a galaxy (SBa) in the ESO/Uppsala Survey of the ESO(B) Atlas \citep{Lau82}, based on observations made with the ESO $1\, \rm{m}$ Schmidt telescope at La Silla, Chile. From its first-ever X-ray observation with \rosat in 1995 and follow-up optical spectroscopy in 1996 with the 2.2m ESO/MPG
telescope at La Silla observatory, \citet{Pie98} have classified it as a Seyfert 1.8 galaxy at a redshift of 0.025701.

The next time \eso was observed in the X-rays was with \xmm in May 2001. From spectral analysis of the 4 ks EPIC-pn data, \citet{Por04} reported the presence of a soft X-ray excess and a highly redshifted \FeKa line at 5.4 keV. The only other X-ray observation of \eso to this date was 100 ks long and performed with \xmm in November 2005. \citet{Por07} reported strong rapid variably from their power spectral density timing analysis of the EPIC-pn data. Furthermore, from spectral analysis of the iron line band, no redshifted \FeKa line was detected contrary to the 2001 findings, while the presence of two narrow emission lines at about 6.5 keV and 7 keV was reported.

In this work we present the first analysis of the RGS spectra from the 100 ks \xmm observation of 2005, which show clear signs of warm absorber outflows. We also used the  simultaneous EPIC-pn and OM data to aid us in the photoionisation modelling of the outflows. The structure of this paper is as follows. Section \ref{obs_sect} describes the observations and data reduction. Section \ref{host_galaxy} focuses on the nuclear obscuration of the AGN. Spectral analysis and photoionisation modelling of the warm absorber are described in Sect. \ref{analysis_sect}. Construction of the SEDs and modelling with different SEDs are described in Sect. \ref{sed_sect}. We discuss our findings in Sect. \ref{discussion} and give concluding remarks in Sect. \ref{conclusions}. 

Note that all the spectra and SEDs displayed in this paper are in the observed frame. The RGS spectra shown are background-subtracted.

\section{Observations and data reduction}
\label{obs_sect}
The ID of the analysed \xmm observation is 0301890101, starting at 22:17 UTC on 10 November 2005 with an EPIC-pn duration of about 102 ks. We extracted the RGS, EPIC-pn and OM data from the \xmm Science Archive. All the observation data files were processed using SAS v10.0.

\subsection{RGS data}
The RGS instruments were operated in the standard Spectro+Q mode. We filtered out time intervals with background count rates $> 0.1\ \mathrm{ct\ s}^{-1}$ in CCD number 9, which is the most affected by background flares. We removed about 8 ks, leaving a net exposure time of about 95 ks for each RGS instrument. The data were processed through the {\tt rgsproc} pipeline task; the source and background spectra were extracted and the response matrices were generated. Using the {\tt rgscombine} task, we combined the RGS1 and RGS2 first-order spectra and response matrices in order to achieve better signal-to-noise for the purpose of spectral fitting. The channels in the RGS spectrum were then re-binned by a factor of 3 to improve statistics. Finally, the modelling of the 6--38 $\AA$ part of the RGS spectrum was performed using the SPEX v2.02.04\footnote{http://www.sron.nl/spex} fitting package.

\subsection{EPIC-pn data}
During the \xmm observation, the EPIC-pn camera was operated in the Full-Frame mode with the medium-filter applied. We extracted a single event, high energy light curve, with the $\mathtt{\#XMMEA\_EP}$ selection attribute, in order to create a set of Good Time Intervals (GTI) excluding intervals of flaring particle background with count rates exceeding ${0.4\ \mathrm{ct\ s}^{-1}}$. We then applied the GTI to the production of the spectra, for which X-ray events with the $\mathtt{FLAG==0}$ attribute were selected. The spectra were extracted from a circular region of $35''$ radius centred on the source. The background was extracted from a nearby source-free region of the same size on the same CCD chip. The EPIC-pn data showed no evidence of pile-up, thus single and double events were selected. Response matrices were generated using the SAS tasks $\mathtt{arfgen}$ and $\mathtt{rmfgen}$. The net exposure time of the EPIC-pn data used for spectral analysis was 77 ks.

\subsection{OM data}
The OM data from Image-mode operations were taken with U, UVW1, UVM2 and UVW2 filters. The OM data were processed with the $\mathtt{omichain}$ pipeline. We performed aperture photometry on each image in a fully interactive way using the $\mathtt{omsource}$ program. The OM count rates were extracted from a circle of 12 pixels radius (5.8\arcsec) centred on the source nucleus. The source is point-like within the extraction region in the UV OM data. The background was extracted from a source-free region of the same radius. We then applied all the necessary corrections, i.e. for the point spread function (PSF) and coincidence losses, including time-dependent sensitivity (TDS) corrections.

\section{Intrinsic obscuration of the nuclear source}
\label{host_galaxy}

\subsection{Nuclear obscuration from the Balmer decrement}
\label{Balmer_red_sect}
\eso is classified as a Seyfert 1.8 galaxy (with strong narrow components and very weak but still visible broad components of \Ha$\lambda6563$ and \Hb$\lambda4861$ emission lines) by \citet{Pie98}, following the \citet{Ost89} observational classification of AGN. The \Ha/\Hb line ratio is not one of the properties given in \citet{Pie98}, so we re-analysed their optical spectrum in order to obtain the Balmer decrement. We fitted both the \Ha and \Hb lines with two narrow and broad Gaussian-profile components. The two forbidden lines of [\ion{N}{ii}] $\lambda\lambda 6548, 6583$, which are blended with \Ha, were also modelled with narrow Gaussian-profile components. To calculate the \Ha/\Hb ratio we used the total flux of each line (including both narrow and broad components) since the broad component of the \Hb is very weak and de-blending of the narrow and broad component of the lines increases the uncertainty in the flux measurements. So the Balmer decrement and consequently the reddening calculated here is an average over the NLR and BLR. We find the observed Balmer decrement $\rm{\Ha/\Hb} = 7.96$. The FWHM is about $2000\ \kms$ for the broad components and about 500 \kms for the narrow components of the above lines.

To convert the Balmer decrement into reddening $E(B-V)$, we used the mean extinction curve of \citet{Gas07} obtained for a sample of AGN using HST. Their best determined extinction curves are flatter in the far-UV and are missing the $\lambda2175$ bump compared to the standard Galactic curve (e.g. \citealt{Ost89}, \citealt{Car89}). The relationship $f_{\lambda, \rm{obs}} = f_{\lambda, \rm{int}}\, 10^{-0.4\, A_{\lambda}}$ gives

\begin{equation}
\label{balmer_eq_1}
\log \left( {\frac{{R_{{\rm{obs}}} }}{{R_{{\rm{int}}} }}} \right) =  - 0.4\, (A_{\lambda ,{\rm{H}}\alpha }  - A_{\lambda ,{\rm{H}}\beta } )
\end{equation}
where $R_{\rm{obs}}$ and $R_{\rm{int}}$ are the observed and intrinsic Balmer decrements \Ha/\Hb; $A_{\lambda ,{\rm{H}}\alpha}$ and $A_{\lambda ,{\rm{H}}\beta}$ represent the extinction at the wavelengths of the Balmer lines. Using the parameterisation of the mean extinction curve of \citet{Gas07}, we obtain the colour excess
\begin{equation}
\label{balmer_eq_2}
E(B - V) = 1.790\, \log \left( {\frac{{R_{{\rm{obs}}} }}{{R_{{\rm{int}}} }}} \right)
\end{equation}
The scalar specifying the ratio of total to selective extinction $R_V \equiv A_V/E(B-V)$ was fixed at 3.1.

From theoretical recombination studies, the Balmer decrement $\rm{\Ha/\Hb} = 2.85$ for Case B recombination at a temperature $T=10^4$~K and electron density $n_{\rm{e}}=10^4\ \rm{cm}^{-3}$ (\citealt{Ost06}). Case B refers to the recombination which takes place in a typical nebula with a large enough optical depth that Lyman-line photons are scattered $n$ times and are converted (if $n \ge 3$) into lower-series photons plus either Ly-$\alpha$ or two-continuum photons, so that they cannot escape from the nebula \citep{Ost06}. For the NLR and BLR of AGN, which have higher electron density than a nebula, this value requires modification due to contribution to the \Ha line from collisional excitation. For the NLR of AGN the intrinsic Balmer decrement \Ha/\Hb is about 3.1 \citep{Ost06}. This value however may be different for the BLR as clouds have higher density in this region. \citet{Fer97} have computed the ratio of Balmer lines relative to \Hb for realistic conditions in BLR clouds. For a BLR with $T=10^4$~K and $n_{\rm{e}}=10^{11}\ \rm{cm}^{-3}$, they obtain an intrinsic $\rm{\Ha/\Hb} = 3.28$. From observations, \citet{War88} have measured the average intrinsic \Ha/\Hb in BLR to be 3.5, independent of any atomic physics assumptions. Using a similar approach, \citet{Car04} have also measured the intrinsic \Ha/\Hb to be 3.43. 

So in our calculations we take $R_{\rm{int}} =\,$3.1\,--\,3.5 and $R_{\rm{obs}}$ to be 7.96, which yields intrinsic $E(B-V) =$~0.64\,--\,0.73. We then use the extinction curve of \citet{Gas07} to correct all the optical-UV data for reddening in the host galaxy of the AGN, using $E(B-V) =$~0.64 and $E(B-V) =$~0.73 for two possible scenarios.

\subsection{Nuclear obscuration from the optical-UV continuum}
\label{continuum_red_sect}
Intrinsic reddening will also affect the optical/UV continuum and change the shape of the so-called `big-blue-bump', which was first attributed to thermal emission from the accretion disc by \citet{Shi78}. The optical/UV continuum is AGN-dominated in \eso; for the purpose of our study contaminating emission by the stars in the host galaxy would have a negligible effect on the SED and photoionisation modelling. We corrected the continuum in such a way that (1) the optical slope ($\alpha_{\rm{opt}}$) and (2) the relation between optical-to-X-ray spectral index ($\alpha_{\rm{ox}}$) and the $\log_{10}$ of the monochromatic optical luminosity at 2500~\AA\ ($l_{\rm{opt}}$) are consistent with AGN samples found in surveys.

\citet{Youn10} have cross-correlated the Sloan Digital Sky Survey (SDSS) DR5 quasar catalog with the \xmm archive, and have obtained $\alpha_{\rm{opt}}$ and $\alpha_{\rm{ox}}$ for a sample of 327 quasars with high X-ray signal-to-noise ratio, where both optical and X-ray spectra are available. \citet{Youn10} find $\alpha_{\rm{opt}} = -0.40$, which is also similar to the slope $\alpha_{\rm{opt}} = -0.44$ found by \citet{Vand01} for mean composite quasar spectra using a dataset of over 2200 spectra from the SDSS. For the uncorrected continuum, we find $\alpha_{\rm{opt}} = -1.74$, which indicates reddening. So, we used the extinction curve of \citet{Gas07} to correct the optical/UV data for reddening to match the $\alpha_{\rm{opt}} = -0.40$. We find that the amount of reddening required for this correction corresponds to $E(B-V) =0.39$, which is less than $E(B-V)$ range of 0.64 to 0.73, inferred from the Balmer decrement.

Furthermore, we checked our values of $\alpha_{\rm{ox}}$ and $l_{\rm{opt}}$ against the $\alpha_{\rm{ox}}$-$l_{\rm{opt}}$ relation found by \citet{Youn10}. For the uncorrected case we find $\alpha_{\rm{ox}} = -1.2$ and $\log\, l_{\rm{opt}} = 28.0$ ($\rm{erg}\, \rm{s}^{-1}\, \rm{Hz}^{-1}$), and for the continuum corrected case with $\alpha_{\rm{opt}} = -0.40$, we find $\alpha_{\rm{ox}} = -1.6$ and $\log\, l_{\rm{opt}} = 29.1$ ($\rm{erg}\, \rm{s}^{-1}\, \rm{Hz}^{-1}$). In both cases, the values we find are consistent within the dispersion in the $\alpha_{\rm{ox}}$-$l_{\rm{opt}}$ relation: for $\log\, l_{\rm{opt}} = 28.0$, $\alpha_{\rm{ox}}$ is between $-1.4$ and $-1.1$; and for $\log\, l_{\rm{opt}} = 29.1$, $\alpha_{\rm{ox}}$ is between $-1.6$ and $-1.3$. On the other hand, for the Balmer-decrement corrected case, $\alpha_{\rm{ox}}$ ($-2.0$ to $-1.9$) is not consistent with the $\alpha_{\rm{ox}}$ of $-1.7$ to $-1.3$ from the $\alpha_{\rm{ox}}$-$l_{\rm{opt}}$ relation for $\log\, l_{\rm{opt}}$ between $29.7$ and $29.9$. This may indicate that the Balmer decrement over-estimates the amount of reddening; we return to this issue in Sect. \ref{obsc_discuss}.

\subsection{Intrinsic X-ray obscuration of the nuclear source by cold neutral gas?}
\label{balmer_column_sect}
From the intrinsic reddening $E(B-V)$ values calculated in Sect. \ref{Balmer_red_sect} from the Balmer decrement and in Sect. \ref{continuum_red_sect}, based on the continuum reddening, the associated column density of cold gas in \eso can be estimated. The relationship between hydrogen column density \NH and optical extinction $A_V$ can be written as ${N_{\rm{H}}\ ({\rm{cm}}^{-2}) = a \times 10^{21}\ A_V\ (\rm{mag})}$, where the factor $a$ is reported to be 2.22 in \citet{Gor75}, 1.79 in \citet{Pre95}, 1.89 in \citet{Ost06} and 2.21 in \citet{Guv09}. Therefore, one can calculate a range of values for \NH using the different values of $E(B-V)$ and $a$ given above. This gives {$\NH =\,$(3.5--5.0)\ $\times 10^{21}\ {\rm{cm}}^{-2}$} if using $E(B-V)$ calculated from the Balmer decrement and {$\NH =\,$(2.2--2.7)\ $\times 10^{21}\ {\rm{cm}}^{-2}$} if using $E(B-V)$ calculated from the continuum reddening.

However, analyses of the X-ray spectra show that \eso is not intrinsically absorbed by a large column of neutral gas. \citet{Por04} obtain an upper limit of $8 \times 10^{19}\ {\rm{cm}}^{-2}$ to intrinsic absorption from analysis of the 2001 EPIC-pn spectrum. This is indeed much lower than the predicted values of {(3.5--5.0)\ $\times 10^{21}\ {\rm{cm}}^{-2}$} or {(2.2--2.7)\ $\times 10^{21}\ {\rm{cm}}^{-2}$} calculated above from the intrinsic reddening. \citet{Por07} also mention that during the 2005 \xmm observation, instead of intrinsic neutral absorption expected below about 1 keV, a soft X-ray excess is observed. From analysis of the 2005 EPIC-pn spectrum, we find an upper limit of $8.9 \times 10^{19}\ {\rm{cm}}^{-2}$ (at 99\% confidence level) to the \NH of intrinsic neutral gas. This upper limit is compatible with that found by \citet{Por04} for the 2001 data. We discuss this apparent discrepancy between the optical-UV reddening and X-ray absorption, and its implications, in Sect. \ref{obsc_discuss}.

\section{Photoionisation modelling and spectral analysis of the warm absorber}
\label{analysis_sect}

\begin{figure*}[!]
\centering
\resizebox{0.82\hsize}{!}{\includegraphics[angle=90]{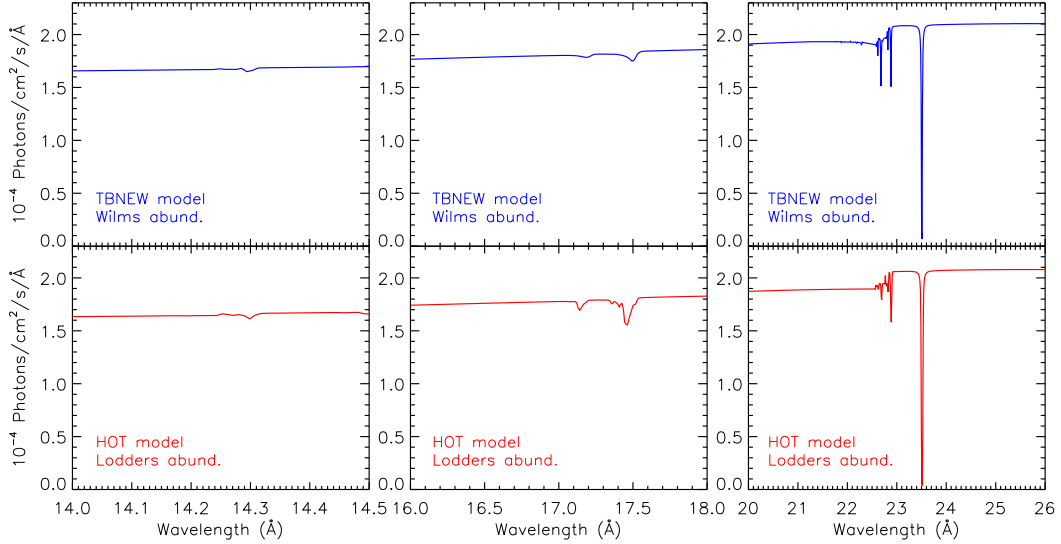}}
\caption{The {\it tbnew} and {\it hot} ISM X-ray absorption models as described in Sect. \ref{analysis_sect} applied to the power-law continuum of \eso, shown in the vicinity of the K-edges of Ne (left panels) and O (right panels) and L-edge of Fe (middle panels).}
\label{tbnew_hot}
\end{figure*}

We began our modelling of the soft X-ray spectrum of \eso by fitting the RGS continuum with a cosmologically redshifted power-law. In all our fits the effects of the Galactic neutral absorption were included by applying the {\it hot} model (collisional ionisation equilibrium) in SPEX. This absorption model uses the \citet{Vern96} cross-sections; for information about the atomic database used in SPEX see the SPEX reference manual. Assuming \citet{Lod09} abundances, the Galactic \ion{H}{i} column density in our line of sight was fixed at $N_{\mathrm{H}}={2.78\times 10^{20}\ \mathrm{cm}^{-2}}$ \citep{Dic90} and the gas temperature at 0.5 eV to mimic a neutral gas. We note here that as requested by the referee of this paper, we also tested the {\it tbnew} model \citep{Wilm11} for X-ray absorption in the ISM using the \citet{Wilm00} abundances. We find the features in the vicinity of the K-edges of Ne and O and L-edge of Fe to be very similar in both the {\it tbnew} and {\it hot} models as shown in Fig. \ref{tbnew_hot}. Therefore, using the {\it tbnew} model would not change the results of our warm absorber study. Furthermore, we checked the impact of using the slightly smaller \NH value of ${2.08\times 10^{20}\ \mathrm{cm}^{-2}}$ from the Leiden/Argentine/Bonn (LAB) Survey of Galactic \ion{H}{i} \citep{Kal05} on our analysis; we find that all parameters derived in this work remain unchanged within the given errors.

The power-law fit, letting photon index and normalisation free, led to a reduced Chi-squared (\redchi) value of 1.6 for 1054 degrees of freedom (d.o.f.). The fit was not satisfactory as there were absorption features, which were not fitted by the Galactic absorption model. At this stage we included one \xabs component in the model. For an assumed SED of the source, \xabs calculates the transmission through a slab of material where all ionic column densities are linked in a physically consistent fashion through a photoionisation model. The ionisation balance calculations were performed using version C08.00 of Cloudy\footnote{http://www.nublado.org} \citep{Fer98} with \citet{Lod09} abundances. For an assumed SED, we made runs with Cloudy for a grid of ionisation parameter ($\xi$) values ($\log \xi$ between $-8.5$ and $+6.5$ with steps of $0.1$) in order to calculate the equilibrium ion concentrations. The ionisation parameter $\xi$ (introduced by \citealt{Tar69}), which is used to describe photoionisation equilibrium, is defined as 
\begin{equation}
\label{xi_eq}
\xi  = \frac{L}{{nr^2 }}
\end{equation}
where $L$ is the luminosity of the ionising source over the 1--1000 Ryd band (in $\rm{erg}\ \rm{s}^{-1}$), $n$ the hydrogen number density (in $\rm{cm}^{-3}$) and $r$ the distance between the ionised gas and the ionising source (in cm). Therefore, $\xi$ is in units of $\rm{erg\ cm}\ \rm{s}^{-1}$. 

Using the auxiliary program {\tt xabsinput} in the SPEX package, the Cloudy output is converted into an input file, containing the temperature and ionic column densities as a function of $\xi$, for the {\it xabs} photoionised absorption model of SPEX. In the modelling we fitted the ionisation parameter ($\xi$), the equivalent hydrogen column density ($\NH$) of the warm absorber, its flow and RMS velocities. As a start, we adopted the SED of the Seyfert-1 galaxy \object{Mrk 509} (but see Sect. \ref{sed_sect} for detailed modelling with different SEDs for \eso). This SED was established with the help of simultaneous data from a large multi-wavelength campaign \citep{Kaa11a}, and broad-band modelling of the underling continuum \citep{Meh11}, and is likely to be representative of the real SED shape for a typical Seyfert galaxy (SED G in Fig. \ref{SED_fig}). The inclusion of the \xabs component improved \redchi to 1.22 (1050 d.o.f.). There were, however, absorption lines in the lower energy part of the spectrum which could not be fitted with a single \xabs phase, since they are less highly ionised and have a different velocity. So we introduced a second \xabs phase in the model to obtain a better fit to the data. The addition of the second phase further improved \redchi to 1.12 (1046 d.o.f.) and all the significantly detected absorption features were fitted. We tested adding another \xabs component, but this did not make the fit any better, so we conclude that two \xabs components are sufficient to model the warm absorption in \eso. Finally, we repeated the above analysis using other SEDs adopted for \eso, which are described in Sec. \ref{sed_sect} and shown in Fig. \ref{SED_fig}.

\begin{figure}[!]
\centering
\resizebox{\hsize}{!}{\includegraphics[angle=0]{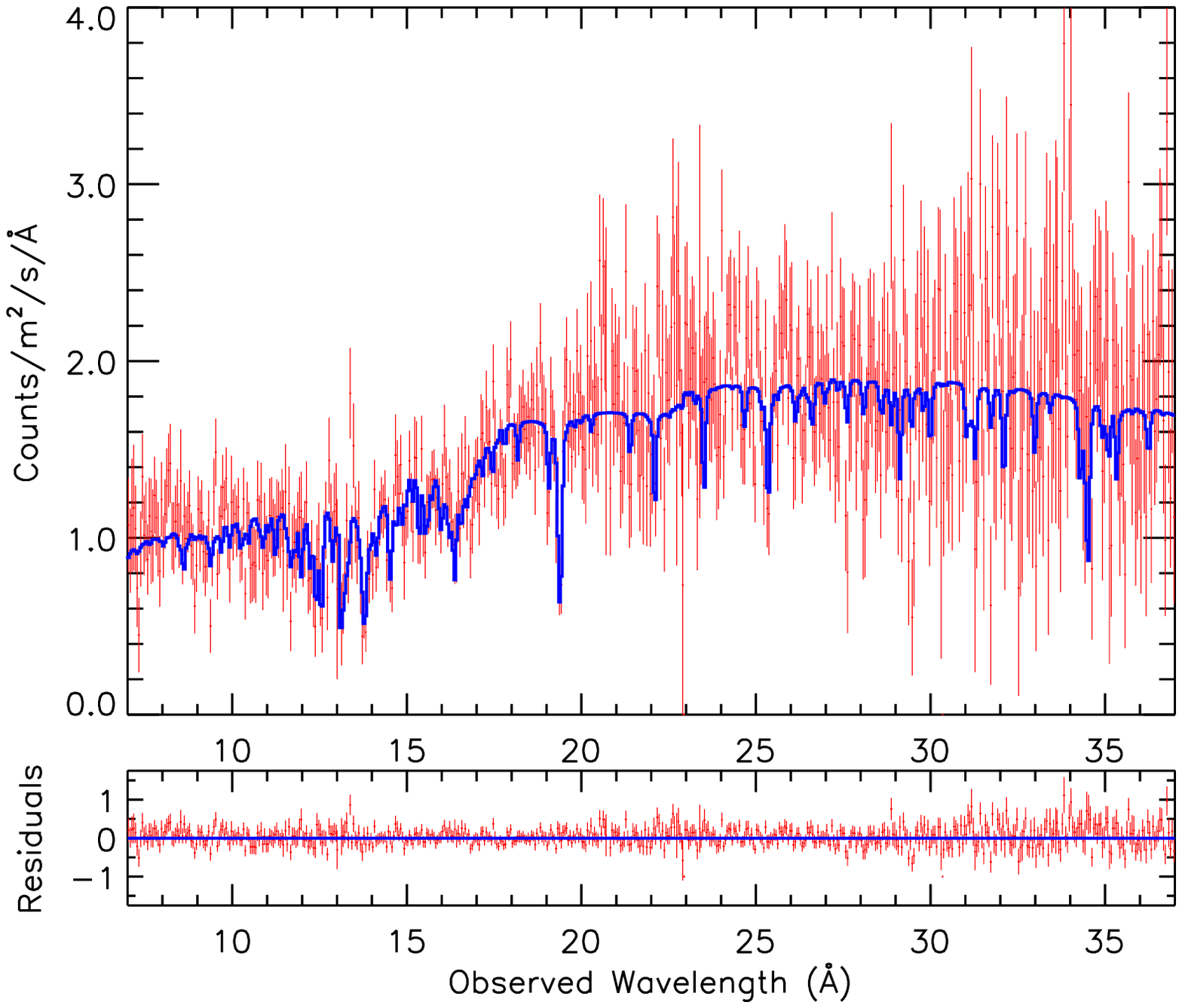}}
\hspace*{-0.4in}
\resizebox{\hsize}{!}{\includegraphics[angle=0]{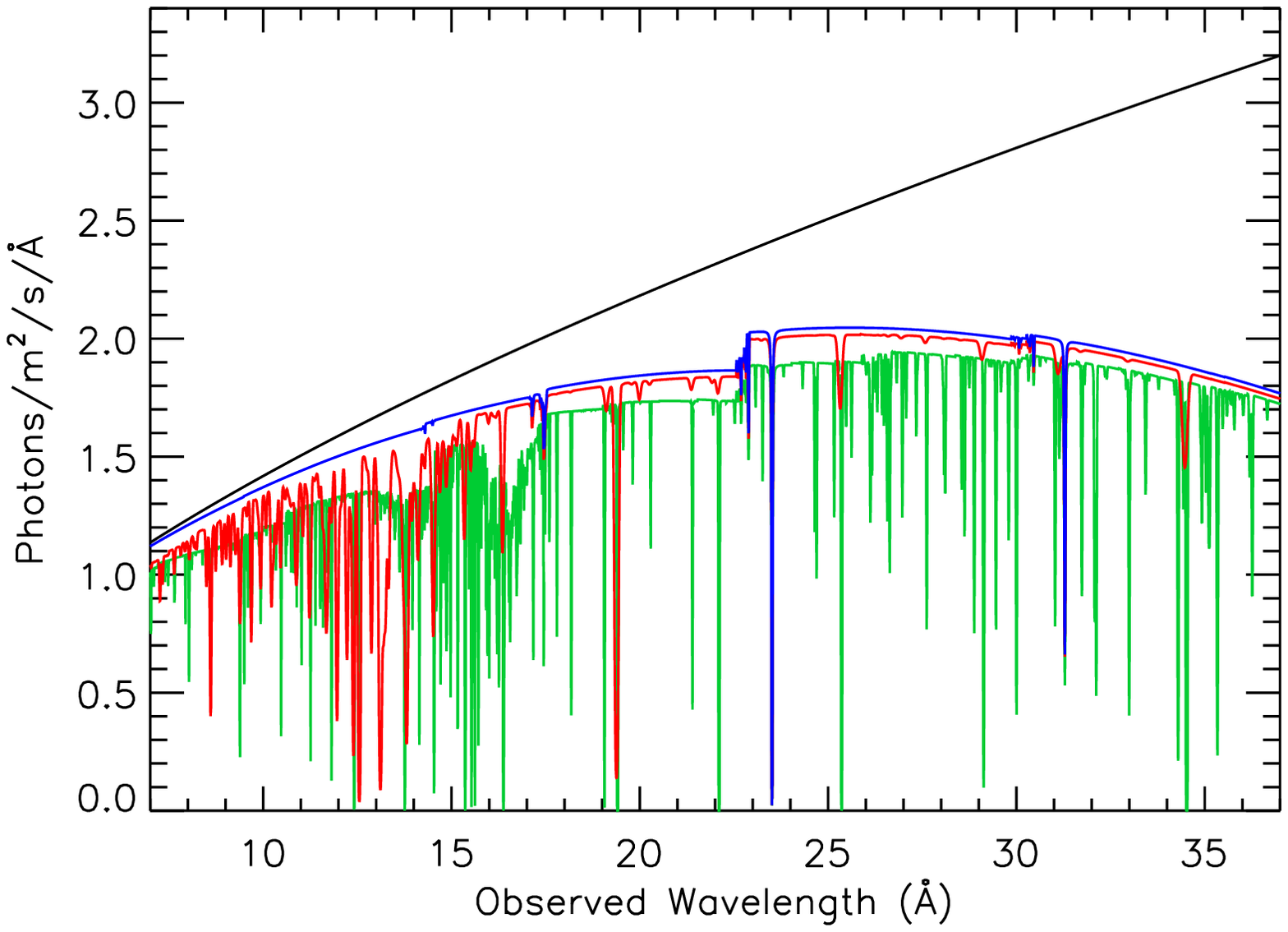}}
\caption{{\it Top panel}: RGS spectrum of \eso, fitted using the two-phase \xabs model described in Sect. \ref{analysis_sect} and SED E in Fig. \ref{SED_fig}. The data are shown in red and the model in blue. Residuals of the fit, (Observed$-$Model)/Model, are displayed below the panel. {\it Bottom panel}: The absorption components of the best-fit RGS model as described in Sect. \ref{analysis_sect}, with the best-fit parameters of the model given in Table \ref{xabs_table}. The blue component shows the Galactic absorption applied to the power-law continuum (in black). The two warm absorber phases (Phase 1 in red and Phase 2 in green) are applied separately to the Galactic-absorbed continuum, showing how each phase contributes to the absorption.}
\label{rgs_spectrum}
\end{figure}

\begin{figure}[!]
\centering
\resizebox{\hsize}{!}{\includegraphics[angle=0]{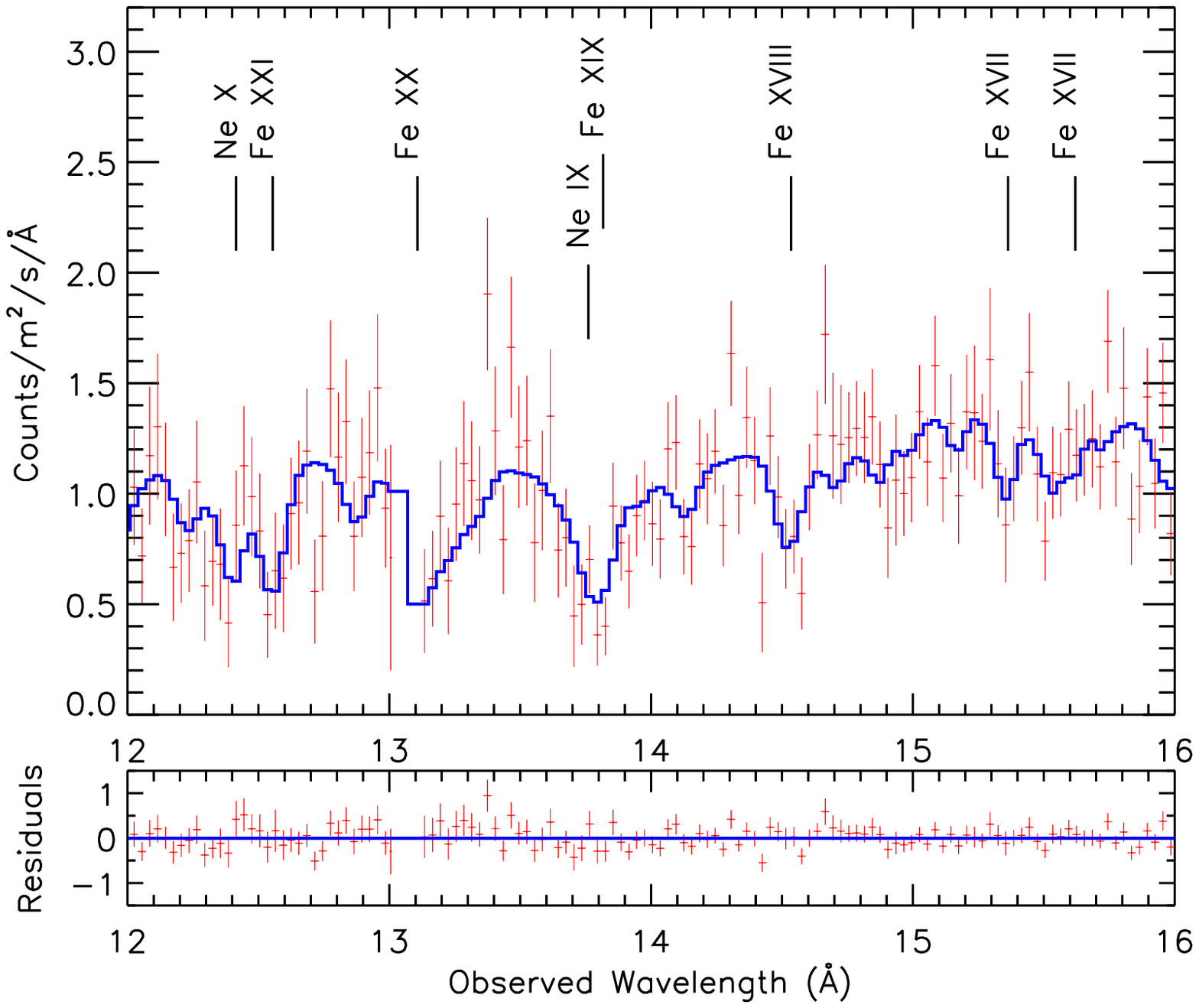}}\vspace*{-0.15in}
\vspace*{-0.15in}
\resizebox{\hsize}{!}{\includegraphics[angle=0]{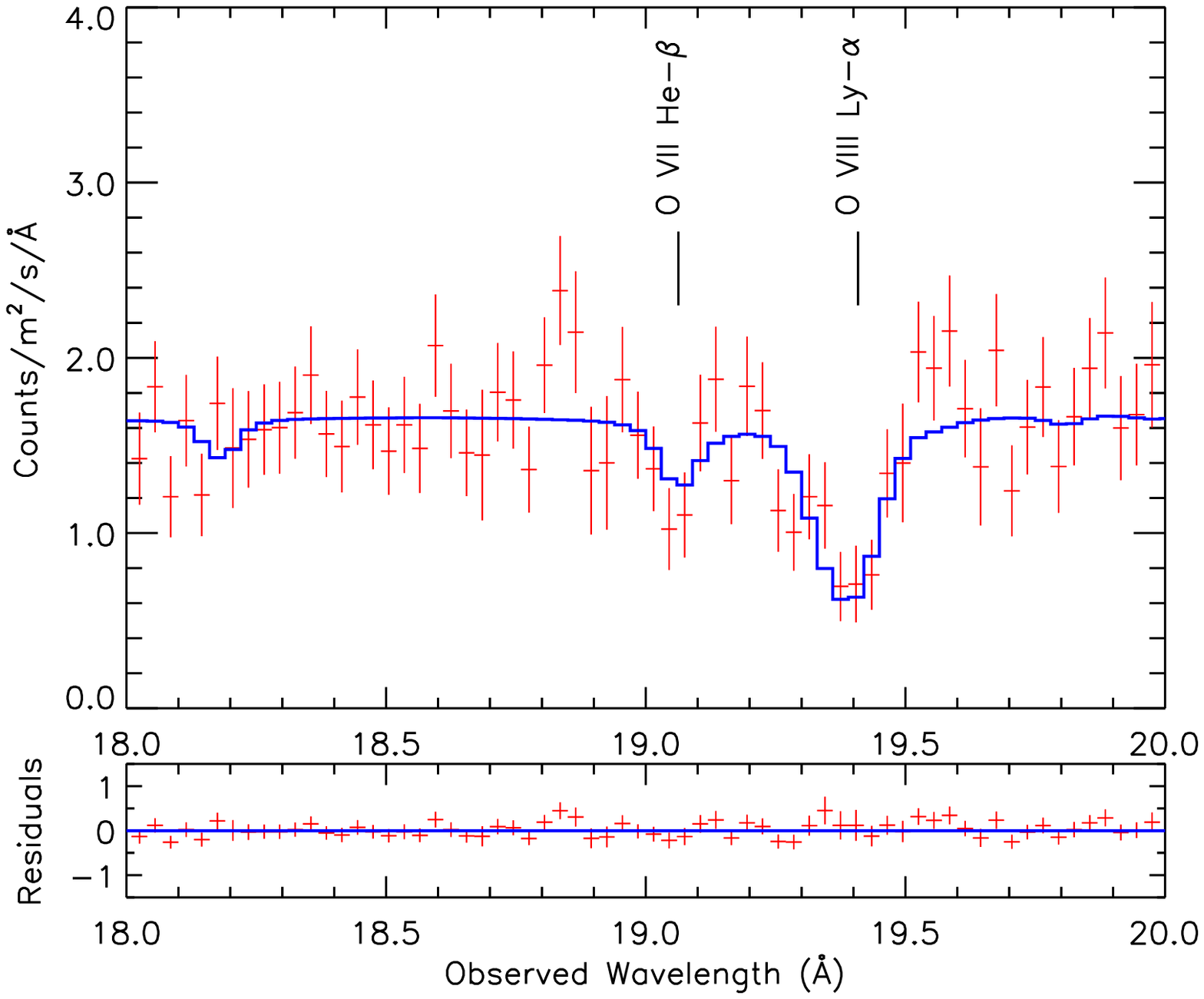}}
\resizebox{\hsize}{!}{\includegraphics[angle=0]{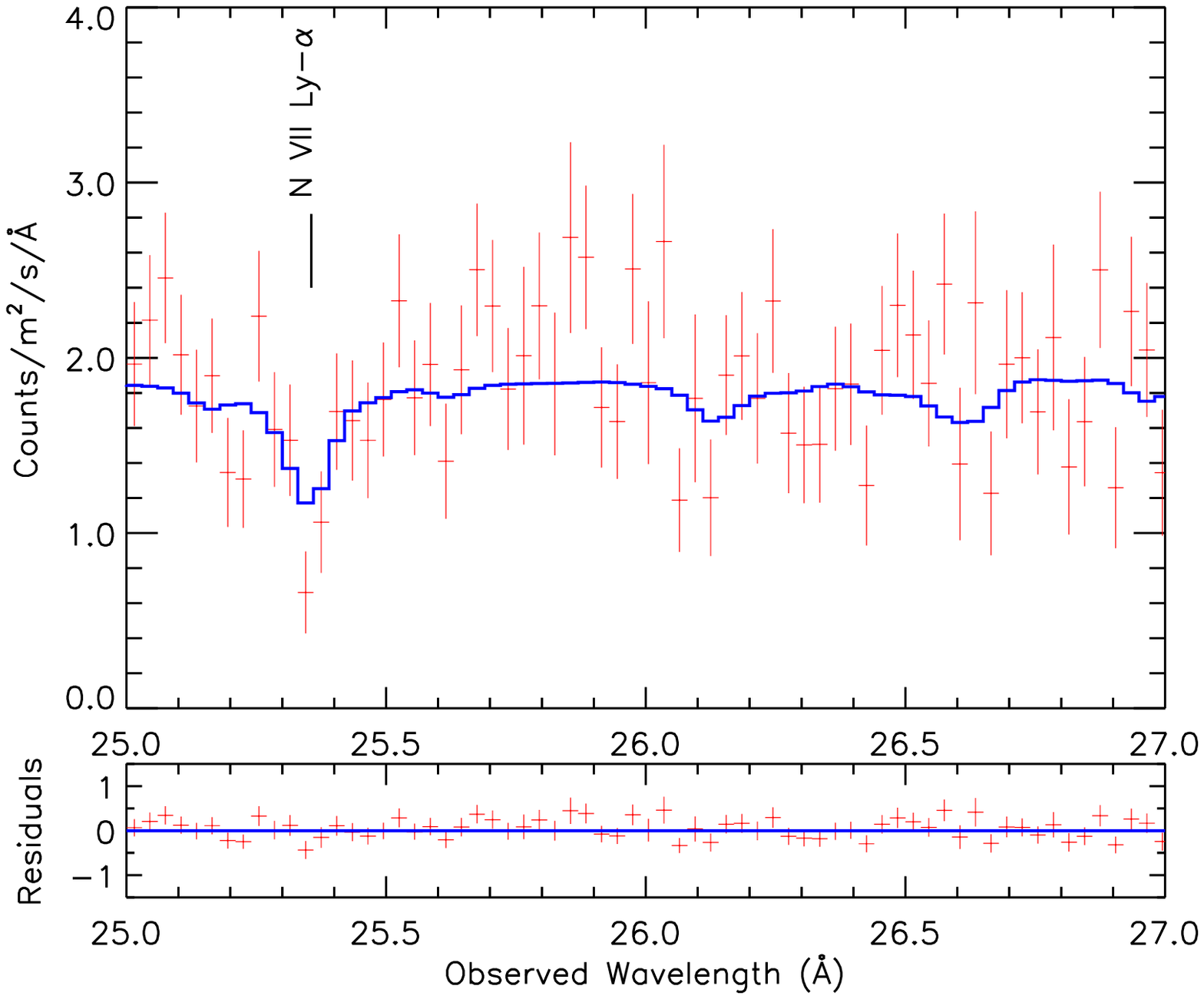}}
\caption{Close-ups of parts of the RGS spectrum and the best-fit model (the two-phase \xabs modelling described in Sect. \ref{analysis_sect} and using SED E in Fig. \ref{SED_fig}) where the most prominent absorption lines are detected. The data are shown in red and the model in blue. Residuals of the fit, (Observed$-$Model)/Model, are displayed below each panel.}
\label{closeups}
\end{figure}

The best-fit parameters of the model (obtained using SED E) are shown in Table \ref{xabs_table}. Note that the velocities of the warm absorber phases are the same for the different SEDs of Fig. \ref{SED_fig}. Furthermore, the only parameter of the warm absorber outflows which significantly changes as a consequence of using different SEDs is the ionisation parameter $\xi$, whose values are given in Fig. \ref{cooling_curve}. The total column densities of different ions in the two warm absorber phases are listed in Table \ref{column_table}. In the top panel of Fig. \ref{rgs_spectrum}, the RGS spectrum and the best-fit model are presented and in the bottom panel we show how each phase in our model contributes to the absorption. In Fig. \ref{closeups}, we show close-ups of the RGS spectrum where the most prominent absorptions lines are detected.

\begin{table}[!]
\caption{Best-fit parameters of the power-law continuum and two-phase warm absorber \xabs model obtained from fitting the RGS spectrum as described in Sect. \ref{analysis_sect} and using SED E in Fig. \ref{SED_fig}. The $\redchi = 1.12$ (1046 d.o.f.). The fitted parameter errors are quoted at 90\% confidence for one interesting parameter.}
\label{xabs_table}
\setlength{\extrarowheight}{3pt}
\begin{minipage}[t]{\hsize}
\renewcommand{\footnoterule}{}
\centering
\begin{tabular}{c c c c c}
\hline \hline
Power-law: & \multicolumn{2}{c}{$\Gamma$ \footnote{Photon index.}} & \multicolumn{2}{c}{Norm \footnote{Normalisation in $10^{51}$ photons $\mathrm{s}^{-1}$ $\mathrm{keV}^{-1}$ at 1 keV.}} \\
 & \multicolumn{2}{c}{$2.62 \pm 0.05$} & \multicolumn{2}{c}{$3.1 \pm 0.1$} \\
\hline
\multicolumn{1}{c}{\xabs} & \multicolumn{1}{c}{$\log \xi$ \footnote{$\mathrm{erg\ cm\ }\mathrm{s}^{-1}$.}} & \multicolumn{1}{c}{$N_{\mathrm{H}}$ \footnote{$10^{22}$ $\mathrm{cm}^{-2}$.}} & \multicolumn{1}{c}{Flow $v$ \footnote{\kms.}} & \multicolumn{1}{c}{RMS $v$ $^e$}  \\
\multicolumn{1}{c}{Phase 1:} & $3.2 \pm 0.1$ & \multicolumn{1}{c}{$1.6 \pm 1.0$} & $-1100 \pm 300$ & $500 \pm 300$ \\
\multicolumn{1}{c}{Phase 2:} & $2.3 \pm 0.1$ & \multicolumn{1}{c}{$0.5 \pm 0.2$} & $-700 \pm 100$ & $100 \pm 40$ \\

\hline
\end{tabular}
\end{minipage}
\end{table}

\begin{table}[!]
\caption{The total column densities of the most relevant ions of the warm absorber outflow, derived from the \xabs modelling described in Sect. \ref{analysis_sect}. Percentages of ionic column density produced by each warm absorber phase are also shown. The ranges of values given correspond to the minimum and maximum values found using the different SEDs of Fig. \ref{SED_fig}.}
\label{column_table}
\setlength{\extrarowheight}{3pt}
\begin{minipage}[t]{\hsize}
\renewcommand{\footnoterule}{}
\centering
\begin{tabular}{l c c c}
\hline \hline
 ion & $\log {N}_{\rm{ion}}$ & Phase 1 & Phase 2\\
 & ($\rm{cm}^{-2}$) & $\%$ & $\%$ \\
\hline
\ion{H}{i}         &  13.33 - 15.30  &  3.22 - 5.76	    &  94.24 - 96.78 \\
\ion{H}{ii}        &	 22.26 - 22.32  &  75.71 - 79.62	    &  20.38 - 24.29 \\
\ion{C}{vi}       &	 17.09 - 17.15  &  7.25 - 9.87           &   90.13 - 92.75 \\
\ion{N}{vii}      &  16.97 - 17.02  &  9.03 - 11.86         &  88.14 - 90.97 \\
\ion{O}{vii}      &	 17.28 - 17.35  &  0.54 - 0.75           &   99.25 - 99.46 \\
\ion{O}{viii}     &  18.11 - 18.16  &  13.28 - 16.95       &   83.05 - 86.72 \\
\ion{Ne}{ix}     &	 17.39 - 17.42  &  2.02 - 2.73           &   97.27 - 97.98 \\
\ion{Ne}{x}      &	 17.63 - 17.66  &  42.76 - 49.58       &   50.42 - 57.24 \\
\ion{Mg}{xi}     &	 17.11 - 17.14  &  12.59 - 17.93       &  82.07 - 87.41 \\
\ion{Mg}{xii}    &	 17.29 - 17.30  &  84.97 - 88.58       &  11.42 - 15.03 \\
\ion{Si}{x}        &	 16.53 - 16.55  &  0.00 - 0.01           &   99.99 - 100.00 \\
\ion{Si}{xi}       &	 16.55 - 16.58  &  0.06 - 0.18           &   99.82 - 99.94 \\
\ion{Si}{xii}      &	 16.45 - 16.50  &  3.23 - 6.25           &  93.75 - 96.77 \\
\ion{Si}{xiii}     &	 17.13 - 17.18  &  61.58 - 70.93       & 29.07 - 38.42 \\
\ion{Si}{xiv}     &	 17.40 - 17.46  &  98.12 - 98.88       & 1.12 - 1.88 \\
\ion{S}{xii}       &	 16.37 - 16.43  &  0.04 - 0.18           &  99.82 - 99.96 \\
\ion{S}{xiii}      & 15.99 - 16.08  &  2.09 - 4.61           &  95.39 - 97.91 \\
\ion{S}{xiv}      & 15.79 - 15.93  &  51.73 - 66.12       & 33.88 - 48.27 \\
\ion{S}{xv}       &	 16.94 - 17.01  &  97.17 - 98.30       & 1.70 - 2.83 \\
\ion{S}{xvi}      &	 16.96 - 17.08  &  99.90 - 99.95       &  0.05 - 0.10 \\
\ion{Fe}{xvii}   &	 16.62 - 16.71  &  2.84 - 10.41         &  89.59 - 97.16 \\
\ion{Fe}{xviii}  & 16.56 - 16.64  &  35.55 - 50.92       &  49.08 - 64.45 \\
\ion{Fe}{xix}    &	 16.81 - 16.85  &  91.08 - 94.10       &  5.90 - 8.92 \\
\ion{Fe}{xx}     &	 17.05 - 17.10  &  99.66 - 99.80       &  0.20 - 0.34 \\
\ion{Fe}{xxi}    &	 17.10 - 17.13  &  99.99 - 99.99       &  0.01 - 0.01 \\
\ion{Fe}{xxii}   &	 16.83 - 16.96  &  100.00 - 100.00   &  0.00 - 0.00 \\
\ion{Fe}{xxiii}  &	 16.42 - 16.71  &  100.00 - 100.00   &  0.00 - 0.00 \\

\hline
\end{tabular}
\end{minipage}
\end{table}

\section{Modelling with different Spectral Energy Distributions}
\label{sed_sect}
In order to calculate a more accurate ionisation balance required to improve the photoionisation modelling of the warm absorber, it is essential to determine the broad-band continuum of the source. Note that there are few flux measurements for \eso compared to some of the well-studied Seyfert galaxies. In the following subsections we describe how different parts of the SED were constructed. In Sect. \ref{alt_sed_sect} we discuss the uncertainties in the construction of the SED due to nuclear obscuration of the source and how, in order to investigate the effects of these uncertainties, we select different SEDs in our further modelling.

\subsection{Interstellar de-reddening of our Galaxy}
\label{Gal_deredd}
We corrected all the optical-UV fluxes for interstellar reddening in our Galaxy using the reddening curve of \citet{Car89}, including the update for near-UV given by \citet{ODo94}. The Galactic interstellar colour excess $E(B-V) = 0.025\ \mathrm{mag}$ is based on calculations of \citet{Sch98} as shown in the NASA/IPAC Extragalactic Database (NED). Also in this case $R_V$ was fixed at 3.1.

\subsection{Optical-IR-Radio part of the SED}
\label{ir_optical_sect}
Optical flux measurements in Cousin's B ($440\ \rm{nm}$) and R ($640\ \rm{nm}$) bands were taken from the Surface Photometry Catalogue of the ESO-Uppsala Galaxies \citep{lau89}. We extracted IR flux measurements at J ($1.25\ \mu \rm{m}$), H ($1.65\ \mu \rm{m}$) and $\rm{K}_{S}$ ($2.17\ \mu \rm{m}$) bands from the 2MASS All-Sky Catalog of Point Sources \citep{Cut03}. Furthermore, we used far-IR data at $12\ \mu \rm{m}$, $25\ \mu \rm{m}$, $60\ \mu \rm{m}$ and $100\ \mu \rm{m}$ from the IRAS Faint Source Catalog \citep{Mos90}, and at $160\ \mu \rm{m}$ from the AKARI/FIS All-Sky Survey Bright Source Catalog \citep{Yam10}. These data points are shown as circles in the SEDs displayed in Fig. \ref{SED_fig}; the red circles are for fluxes only corrected for Galactic extinction as described in Sect. \ref{Gal_deredd}, whereas the fluxes shown as blue circles are also corrected for extinction by the host galaxy of the AGN as described in Sects. \ref{Balmer_red_sect} (SEDs C and D) and \ref{continuum_red_sect} (SEDs E and F). We note that the fluxes used here are AGN-dominated but include some stellar optical emission coming from the nuclear region of the galaxy and IR dust emission from the whole galaxy; however, for the purpose of our study, contaminating emission by the host galaxy would have a negligible effect on the broad-band SED and photoionisation modelling.

The only radio-band flux measurement available for \eso was found in the Sydney University Molonglo Sky Survey (SUMSS) Source Catalog \citep{Mau03}. The flux at $36\ \rm{cm}$ (outside the plots in Fig. \ref{SED_fig}) is $13.9 \pm 1.0\ \rm{mJy}$.

\subsection{UV-X-ray part of the SED}
\label{uv_xray_sect}
\begin{figure*}
\centering
\resizebox{0.92\hsize}{!}{\includegraphics[angle=0]{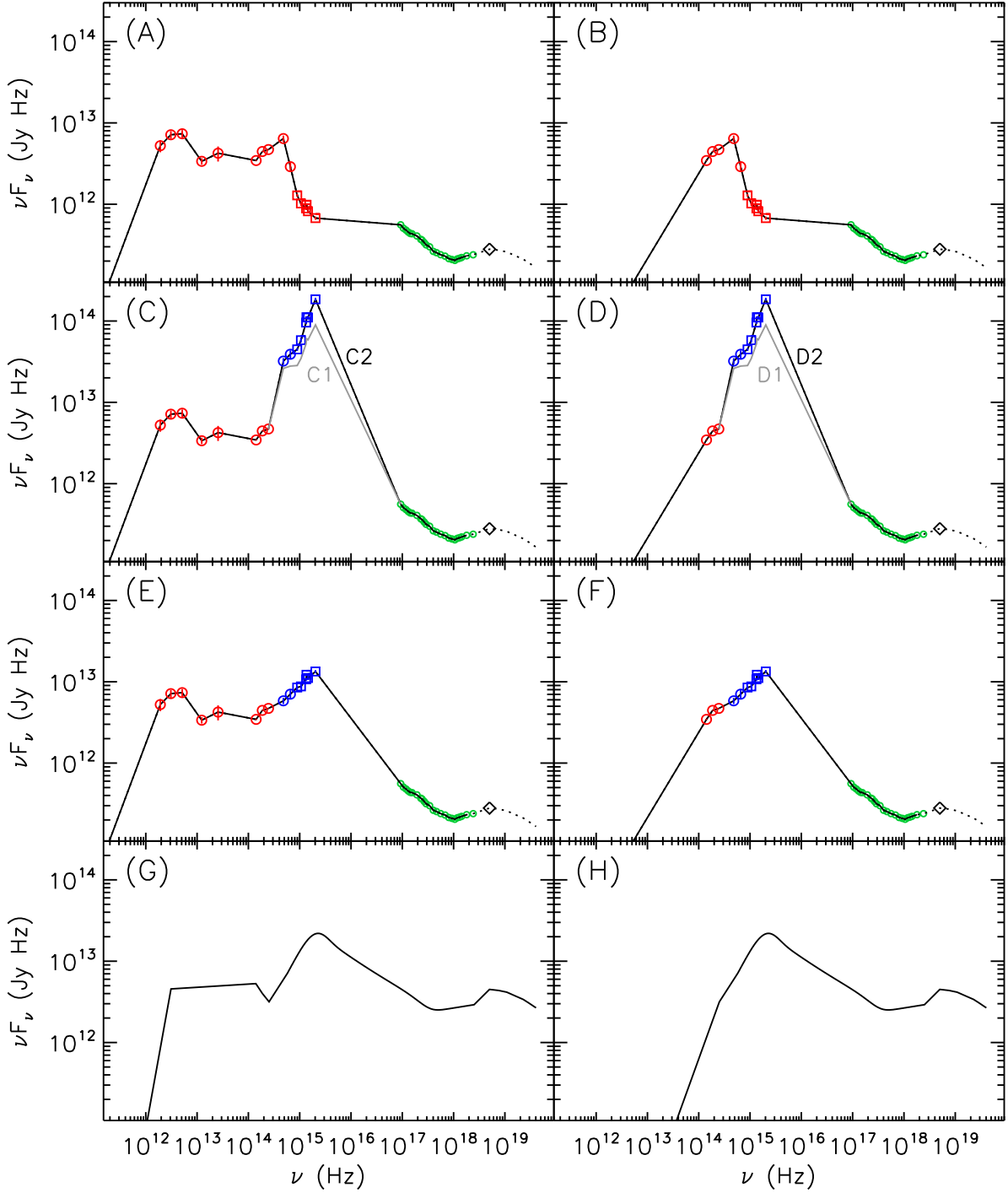}}
\caption{The different SEDs (A--H) assumed in the modelling of \eso warm absorber as described in Sects. \ref{analysis_sect} and \ref{sed_sect}, which were used in the ionisation balance calculations. The IR and optical data points (Sect. \ref{ir_optical_sect}) are shown as circles and the UV data (Sect. \ref{uv_xray_sect}) as squares. The data points in red are corrected only for Galactic extinction as described in Sect. \ref{Gal_deredd}, whereas the blue data points (in panels C, D, E and F) are also corrected for nuclear obscuration in the AGN as described in Sects. \ref{Balmer_red_sect} and \ref{continuum_red_sect}. The SEDs in panels C and D have been corrected for nuclear obscuration by using the Balmer decrement and the ones in panels E and F have been corrected for nuclear obscuration by using the continuum reddening. In panels C and D, the SEDs with some parts shown in grey and no data points correspond to nuclear obscuration correction with an intrinsic $E(B-V)=0.64$ (named SEDs C1 and D1) and the ones with blue data points superimposed on the curve correspond to nuclear obscuration correction with an intrinsic $E(B-V)=0.73$ (named SEDs C2 and D2). The radio point (Sect. \ref{ir_optical_sect}) is not displayed in this figure due to its low flux. The EPIC-pn X-ray data corrected for Galactic absorption are shown as small green circles. The black diamonds correspond to flux at 20 keV extrapolated from the EPIC-pn spectrum, and the extrapolations to the higher energy parts of the SED, shown as dotted lines, are described in Sect. \ref{uv_xray_sect}. SED G is the SED of Mrk~509 introduced in Sect. \ref{analysis_sect} and used as a `standard' representation for a Seyfert AGN. SEDs on the right-hand side (B, D, F, H) are the same as their adjacent SEDs on the left (A, C, E, G) but without the IR bump and are used in modelling cases where the warm absorber does not receive IR radiation from the dusty torus.}
\label{SED_fig}
\end{figure*}

The UV part of the SED was constructed from the \xmm OM observation and the GALEX All-sky Imaging Survey (AIS). The 2005 OM data were taken in the U ($3440\ \AA$), UVW1 ($2910\ \AA$), UVM2 ($2310\ \AA$) and UVW2 ($2120\ \AA$) filters. The 2007 GALEX data were taken in the NUV ($2267\ \AA$) and FUV ($1516\ \AA$) filters. Note that the OM UV data (which are simultaneous with the X-rays) are not simultaneous with the GALEX UV data. However, the OM UVM2 flux is consistent with the GALEX NUV flux, which indicates that the UV flux was at a similar level during the \xmm and GALEX observations. The OM and GALEX data are shown as squares in the SEDs displayed in Fig. \ref{SED_fig}; the red squares show fluxes only corrected for Galactic extinction as described in Sect. \ref{Gal_deredd}, whereas the blue squares are for fluxes also corrected for extinction in the host galaxy of the AGN as described in Sects. \ref{Balmer_red_sect} (SEDs C and D) and \ref{continuum_red_sect} (SEDs E and F).

The 0.3--10 keV X-ray part of the SED was constructed using the EPIC-pn data, corrected for the Galactic \ion{H}{i} column density in our line of sight, $N_{\mathrm{H}}={2.78\times 10^{20}\ \mathrm{cm}^{-2}}$, given by \citet{Dic90}. The EPIC-pn data are shown as small green circles in the SEDs displayed in Fig. \ref{SED_fig}. Unfortunately, there are no X-ray data available for \eso above 10 keV; therefore in order to estimate the SED above 10 keV, we extrapolated the high energy part of the EPIC-pn spectrum using a power-law model up to 20 keV ($4.84\times10^{18}$ Hz); then from 20 keV to 162 keV, the INTEGRAL data from the SED of Mrk 509 as given in \citet{Kaa11a} were used, but scaled down to match the \eso flux at 20 keV. This way the shape of the \eso SED above 10 keV resembles the Compton `hump' commonly seen in Seyfert AGN. We have tested whether the shape of the continuum above 10 keV, and in particular the shape and flux of the Compton hump, affect the results of the warm absorber analysis and the thermal stability curves (Sect. \ref{structure_sect}); we find that changing the hard X-ray continuum does not modify our results and all the derived parameters remain unchanged within the given errors.

\subsection{Different assumed SEDs for modelling}
\label{alt_sed_sect}

Different SEDs of the ionising source can have a significant effect on the ionisation balance of the absorbing material and consequently affect the stability and structure of the warm absorber. So we have taken into account the uncertainties by adopting different SEDs in our further modelling of the warm absorber. 

For the case of \eso, the largest uncertainty arises from the level of extinction and reddening in the host galaxy of the AGN, reported in Sect. \ref{host_galaxy}: (1) there is uncertainty in measuring the Balmer decrement in \eso and (2) the Balmer decrement can be intrinsic to the BLR or NLR clouds, so there are uncertainties on deriving reddening information from the Balmer decrement. As shown in Fig. \ref{SED_fig}, we considered SEDs corrected only for Galactic extinction (SEDs A, B), and with also the AGN obscuration taken into account as described in Sect. \ref{Balmer_red_sect} (SEDs C, D) based on the Balmer decrement and Sect. \ref{continuum_red_sect} (SEDs E, F) based on the continuum reddening. Furthermore, the precise locations of the warm absorber outflows relative to the putative dusty torus are unknown. A dusty torus is expected to exist around the nuclear source in AGN and be responsible for the IR `bump' seen in their SEDs by reprocessing of the nuclear high energy emission. So there is uncertainty as to how much IR radiation is seen by the warm absorber outflows. In addition, \citet{Por07} report that the 2001 EPIC-pn data show no significant constant narrow 6.4 keV \FeKa line (with an upper limit of EW $<$ 32 eV), hence suggesting lack of any dominant Compton reflection emission from distant cold matter such as the dusty torus. Thus, to investigate the effect of emission from the dusty torus in our ionisation balance calculations, we consider the same SEDs as described above but without the IR bump (SEDs B, D, F, H in Fig. \ref{SED_fig}) to represent cases in which the outflow does not see emission from the dusty torus.

We adopted these SEDs in our calculations of the ionisation balance and thermal stability curves, as described in Sect. \ref{structure_sect}. We note here that adopting different SEDs does not affect the goodness of the RGS fits since the columns of the relevant ions detected by the RGS are not significantly different for different SEDs. The parameter which changes significantly as a result of using different SEDs is the ionisation parameter $\xi$, which is given in Sect. \ref{structure_sect} (Fig. \ref{cooling_curve}) for each case and plays a significant role in determining the structure of the warm absorber using thermal stability curves.

\subsection{Different thermal stability curves for the warm absorber}
\label{structure_sect}
The ionisation balance required for the photoionisation modelling is determined by the SED. We now study the structure of the warm absorber for each corresponding SED case in Fig. \ref{SED_fig}. To investigate the stability of each warm absorber phase and the possibility of the two phases co-existing in pressure equilibrium, the pressure form of the ionisation parameter, $\Xi$ (introduced by \citealt{Kro81}), needs to be used. The parameter $\Xi$ is defined as
\begin{equation}
\label{big_xi_eq}
\Xi  = \frac{L}{{4\pi cr^2 nkT}}
\end{equation}
where $c$ is the speed of light, $k$ the Boltzmann constant and $T$ the gas temperature (and $L$, $r$ and $n$ as in Eq. \ref{xi_eq}). Substituting Eq. (\ref{xi_eq}) into Eq. (\ref{big_xi_eq}) gives
\begin{equation}
\label{big_xi_eq_v2}
\Xi  = \frac{\xi }{{4\pi ckT}} \approx 19222\ \frac{\xi }{T}
\end{equation}

\begin{figure*}[!]
\centering
\resizebox{0.93\hsize}{!}{\includegraphics[angle=0]{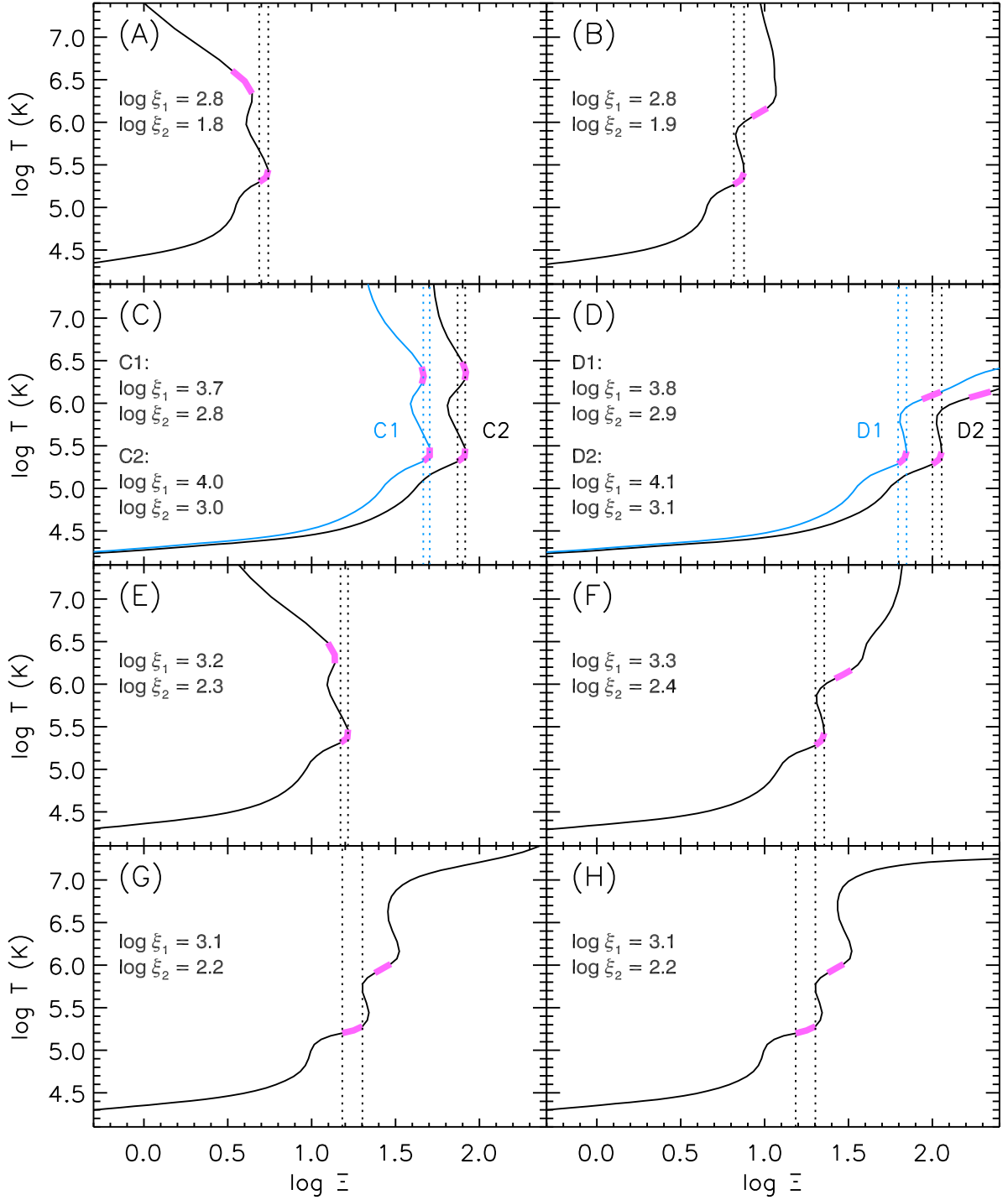}}
\caption{Different stability curves (A--H) calculated for the corresponding SEDs (A--H) of Fig. \ref{SED_fig}. In panels C and D, the curves in blue correspond to SEDs C1 and D1 and the ones in black to SEDs C2 and D2. The positions of the warm absorber phases (Phase 1 and 2), along with their $2\sigma$ (95.4\%) errors, are indicated as thick magenta strips on the curves. Phase 1 is the one with higher $T$. The ionisation parameters of Phase 1 and 2 (i.e. $\xi_{1}$ and $\xi_{2}$) in units of $\mathrm{erg\ cm\ }\mathrm{s}^{-1}$ are given for each corresponding stability curve. The dotted lines are plotted for display purposes to show the range of $\Xi$ for one of the phases on each curve. The regions overlapping in $\Xi$ on each stability curve are in pressure equilibrium.}
\label{cooling_curve}
\end{figure*}

For each SED we produced the corresponding thermal stability curve, shown in Fig. \ref{cooling_curve}, using the output of the Cloudy runs (see Sect. \ref{analysis_sect}). The best-fit ionisation parameters of Phase 1 and Phase 2 and their errors are marked on the stability curves. In order for the two phases to co-exist in pressure equilibrium, they must have overlapping values of $\Xi$. We discuss these results in Sect. \ref{struc_discussion}.

\section{Discussion}
\label{discussion}

\subsection{Nuclear obscuration or an intrinsically large Balmer decrement?}
\label{obsc_discuss}

Here we discuss several possibilities for the discrepancy between the large Balmer decrement in \eso and the fact that the X-ray spectrum is not intrinsically absorbed by cold neutral gas. We note that \citet{Por07} obtained an upper-limit to the neutral gas column density in \eso from analysis of the EPIC-pn spectrum, which is much lower than that predicted for a typical Seyfert 1.8 galaxy, in which the BLR is expected to be reddened. \citet{Por07} briefly mentioned possible explanations for such a discrepancy (explanations 1, 2 and 4 which we expand on below), but did not go into details. In this work, with the benefit of our simultaneous optical/UV (OM) and X-ray (RGS, EPIC-pn) analysis of the \xmm data and also analysis of the archival optical spectrum of \eso, we can discuss the likelihood of each explanation and suggest which is the most viable solution.

\subsubsection*{(1) Non-simultaneous optical and X-ray/UV observations}

The optical spectrum in which \eso shows the large Balmer decrement was taken in 1996, whereas the \xmm observations were taken in 2001 and 2005. The nearest X-ray data to the time of the optical observation were obtained in 1995 by ROSAT. So we can compare the X-ray flux in 2001 and 2005 with the flux in 1995 to check the possibility of variability. As given in \citet{Pie98}, the soft X-ray (0.1--2.4 keV) energy flux from the 1995 ROSAT observation is ${8.5\times 10^{-12}\ \rm{erg\ s}^{-1}\, \rm{cm}^{-2}}$. We calculate the soft X-ray flux in the same band to be ${3.7\times 10^{-12}\ \rm{erg\ s}^{-1}\, \rm{cm}^{-2}}$ in 2001 and ${5.8\times 10^{-12}\ \rm{erg\ s}^{-1}\, \rm{cm}^{-2}}$ in 2005 from the \xmm observations. This flux change over the years is too small to be attributed to occultation by the column of gas inferred from the Balmer decrement in Sect. \ref{balmer_column_sect}: {$\NH =\,$(3.5--5.0)\ $\times 10^{21}\ {\rm{cm}}^{-2}$}. Assuming such a column density of cold gas was present during the optical observation and the unabsorbed soft X-ray flux was at the same level as in 2001 or 2005 (when we know there was no intrinsic neutral absorption) would imply an obscured soft X-ray flux smaller by a factor of 3.1--4.1, whereas in 1995 the X-ray flux was actually larger by a factor of 1.5--2.3.

More importantly, during the 2005 \xmm observation, when simultaneous UV (4 UV filters of OM) and X-ray observations are available, the optical-UV continuum, which is supposed to represent thermal emission from the accretion disc, appears to be intrinsically reddened as described in Sect. \ref{continuum_red_sect}. So the non-simultaneity of the optical spectroscopy and X-ray observations is unlikely to be the cause of discrepancy between the reddening and X-ray neutral absorption.

\subsubsection*{(2) Partially covering neutral gas clouds}
The clouds of neutral gas and dust that obscure the NLR and BLR may only partially cover the nuclear source, such that they are not in our line of sight to the X-ray emitting region. This way optical emission from the NLR and BLR is reddened but the X-rays remain unabsorbed. This scenario is however an unlikely possibility as it requires an ad hoc geometry of the absorbing regions. Furthermore, since the slope of the optical/UV continuum shows that intrinsic UV emission from the accretion disc has been reddened, the intrinsic X-ray emission originating from the corona in close vicinity of the disc must also have passed through the same material as the UV emission. Therefore, partially covering neutral gas clouds in the host galaxy of the AGN are not a feasible solution.

\subsubsection*{(3) Intrinsically large Balmer decrements}

One possible explanation is that Balmer decrements are intrinsically large for the BLR or even the NLR. The physical conditions in the BLR are to some extent uncertain and higher density clouds within the BLR can cause collisional and radiative effects which can raise the value of the intrinsic Balmer decrement, e.g. from 3 to 10 \citep{Kwa81}.

\citet{Bar03} have analysed the optical (WHT) and X-ray (\xmm EPIC-pn) spectra of the Seyfert 1.8/1.9 galaxy \object{H1320+551}. They find that the Balmer decrement \Ha/\Hb is about 6 for the NLR and about 27 or more for the BLR. Despite such large Balmer decrements, the X-ray spectrum is not intrinsically absorbed. They find a $3\sigma$ upper limit of $1.4 \times 10^{20}\ {\rm{cm}}^{-2}$ to the obscuring column density, which they report is about 70 and 7 times smaller than the minimum predicted from the BLR and NLR Balmer decrements respectively. \citet{Bar03} also rule out the existence of an ionised absorber in H1320+551 from their X-ray spectral fits. They suggest the large Balmer decrement of the BLR is an intrinsic property and not caused by internal reddening, and so H1320+551 is not consistent with being an obscured Type-1 Seyfert AGN. 

However, in \eso, the UV continuum provides an independent estimate of the reddening from that inferred from the Balmer decrement. As shown in Sect. \ref{continuum_red_sect}, the amount of reddening derived from the continuum study, $E(B-V) =0.39$, is less than that inferred from the Balmer decrement, $E(B-V) =$~0.64\,--\,0.73. This may indicate that the Balmer decrement over-estimates the amount of reddening, as the intrinsic $\rm{\Ha/\Hb}$ is larger than the theoretical values 3.1\,--\,3.5. From Eq. \ref{balmer_eq_2}, the intrinsic $\rm{\Ha/\Hb}$ needs to be about 4.8 in order to imply $E(B-V) =0.39$, while the value observed is $\sim 8$. One reason disfavouring the large reddening obtained from the Balmer decrement is that the bolometric luminosity calculated from the de-reddened SEDs (SEDs C1 and C2 in Fig. \ref{SED_fig}) is high for a Seyfert galaxy: 2.8--5.8 $\times 10^{46}\ \rm{erg\ s}^{-1}$; on the other hand, the bolometric luminosity of the SED de-reddened using the continuum (SED E in Fig. \ref{SED_fig}) is $4.7 \times 10^{45}\ \rm{erg\ s}^{-1}$. This is very similar to the bolometric luminosity of Mrk 509, calculated from the SED G: $4.6 \times 10^{45}\ \rm{erg\ s}^{-1}$. Nonetheless, even if to some degree the large Balmer decrement is intrinsic, the slope of the optical-UV continuum indicates there must be intrinsic reddening taking place in our line of sight in \eso.

\subsubsection*{(4) Dusty warm absorber}
The most viable explanation for the discrepancy between the reddening and X-ray absorption in \eso is that the X-ray warm absorber contains dust. The dust causes the observed optical-UV reddening of the continuum and the large Balmer decrement, whilst the X-rays remain photoelectrically unabsorbed since the gas in the warm absorber is ionised (predominantly due to the ionisation of carbon and oxygen), and the X-ray opacity of dust is lower than that of cold neutral gas. This means unusually large dust-to-gas ratio clouds in the ISM of the host galaxy are not needed to explain the discrepancy between reddening and X-ray absorption. Previous studies show that dust particles can survive the dust destruction mechanisms (sublimation and thermal sputtering) under conditions found in some warm absorbers. For example, \citet{Rey97} suggest the existence of a dusty warm absorber in the Seyfert-1 galaxy \object{MCG--6-30-15}, to account for the discrepancy between the reddening and lack of X-ray absorption by neutral gas, and discuss the survival of dust grains under conditions found in a photoionised warm absorber. We check here if dust can survive in the warm absorber of \eso.

As shown in \citet{Barv87}, the sublimation radius (the minimum distance from the central source at which particular grains can exist) for graphite grains is given by 
\begin{equation}
R_{\rm{sub}}= 1.3\, L_{\rm{uv,46}}^{0.5}\, T_{1500}^{-2.8}\ \rm{pc}, 
\end{equation}
where $L_{\rm{uv,46}}$ is the UV luminosity of the central source in units of $10^{46}\ \rm{erg}\ \rm{s}^{-1}$, and $T_{1500}$ is the graphite grain sublimation temperature in units of $1500\ \rm{K}$. Taking an upper estimate of $10^{46}\ \rm{erg}\ \rm{s}^{-1}$ for the UV luminosity of \eso (consistent with our SEDs), we obtain ${R_{\rm{sub}} \sim 1.3\ \rm{pc}}$. So as long as the dust resides in a warm absorber phase which is further than ${1.3\ \rm{pc}}$ from the source, it is not destroyed by sublimation.

Another mechanism that can destroy dust grains is sputtering, in which atoms or molecules are knocked off the surface of the dust grain due to collisions with hot ions. As given in \citet{Burk74}, the threshold for sputtering graphite grains corresponds to a gas temperature of about ${4 \times 10^{5}\ \rm{K}}$. \citet{Burk74} have shown that above the sputtering threshold, the grain lifetime against destruction by sputtering is 
\begin{equation}
t_{\rm{sput}} = 6.25 \times 10^{11} (YnT_{4}^{0.5})^{-1}\ \rm{s},
\label{sput_eq}
\end{equation}
where $Y$ is the sputtering yield at $T_{4}$, the incident particle energy in units of ${10^{4}\ \rm{K}}$, and $n$ is the incident particle density. Phase 2 of the warm absorber in \eso has a maximum temperature of ${2.6 \times 10^{5}\ \rm{K}}$ corresponding to SED C1 (see Fig. \ref{cooling_curve}). This temperature is below the sputtering threshold and thus graphite dust grains in Phase 2 should not be destroyed by the sputtering mechanism. On the other hand, Phase 1, which is hotter than Phase 2, has a minimum temperature of ${9.3 \times 10^{5}\ \rm{K}}$ (corresponding to SED G) and a maximum temperature of ${2.7 \times 10^{6}\ \rm{K}}$ (corresponding to SED A). So the temperature of Phase 1 is higher than the sputtering threshold for all SED cases. Substituting $n$ from the expression of the ionisation parameter (${\xi  = L/nr^{2}}$) into that of $t_{\rm{sput}}$ (Eq. \ref{sput_eq}), we find that for Phase 1 (with the parameter values given in Table \ref{xabs_table}), ${t_{\rm{sput}} \sim 22.8 \, r^{2}\ \rm{yr}}$, where $r$ is the distance of the gas from the central source in pc. Furthermore, we can roughly estimate Phase 1 flow timescale, ${t_{\rm{flow}} \sim r/v}$, where $v$ is the velocity of Phase 1 (i.e. $1100\ \kms$); so ${t_{\rm{flow}} \sim 890\, r\ \rm{yr}}$, where the distance $r$ is in pc. The dust grains are destroyed by sputtering if ${t_{\rm{sput}} < t_{\rm{flow}}}$; therefore, if $r < 39\ \rm{pc}$, then the dust grains in Phase 1 cannot survive the sputtering. Assuming a constant velocity outflow, an upper limit for the distance $r$ can be estimated by using $\NH = n\,r\,f$ to substitute into ${\xi  = L/nr^{2}}$ and eliminate $n$, where we take the volume filling factor $f=1$. For Phase 1 we obtain $r < 52\ \rm{pc}$, so the dust grains can only survive in Phase 1 if $39 < r < 52\ \rm{pc}$. However, Phase 2 (with $r < 1.3\ \rm{kpc}$ applying the same calculations as above), which is the lower-ionised phase of the warm absorber with a temperature below the sputtering threshold, is more likely to host dust grains than Phase 1.

\subsection{The structure of the warm absorber}
\label{struc_discussion}
From modelling of the RGS spectrum, we found that the warm absorber in \eso consists of at least two phases of ionisation. As shown in Table \ref{xabs_table}, the higher-ionised Phase 1 has a larger column density than the lower-ionised Phase 2, and also has larger outflow and RMS velocities. The thermal stability curves of Fig. \ref{cooling_curve} show that depending on which ionising SED is selected, quite different stability curves are obtained. The SED also dictates whether the absorber phases lie on the stable or unstable part of the curves. Only in case C, in which the optical-UV continuum is de-reddened using the Balmer decrement and the warm absorber does receive IR radiation from the dusty torus of the AGN, the two phases have overlapping values of $\Xi$ and hence can be in pressure equilibrium. However, for all other cases, the two phases do not overlap in $\Xi$, so they are not in pressure equilibrium. Since also their velocities are different, we conclude that the two phases are likely to be separated and out of pressure equilibrium.

The exact location of the warm absorber relative to the dusty torus is unknown, so it is not clear whether or not the warm absorber sees IR radiation from the torus. As mentioned in Sect. \ref{alt_sed_sect}, from analysis of the \FeKa line, \citet{Por07} report lack of Compton reflection from distant cold regions such as the dusty torus. However, lack of Compton reflection from cold matter does not imply absence of a dusty torus, because the BLR is likely to be obscured by the torus given the lack of broad optical emission lines in this object; also it does not imply the warm absorber is not receiving IR emission from the torus because the warm absorber phases (most likely evaporated off the torus itself), are likely to be further away from the torus (see the $r$ values derived in the previous section), so they may see IR radiation from the outer parts of the torus. As inferred from Fig. \ref{cooling_curve}, apart from cases C and D, the IR radiation from the dusty torus does not affect whether or not the two phases are in pressure equilibrium anyway.

As discussed earlier, dust can survive in the lower-ionised phase (Phase 2) of the warm absorber, which can account for the discrepancy between the large optical/UV reddening and the low amount of X-ray neutral absorption. Since the NLR is reddened as inferred from the Balmer decrement, then the dust in Phase 2 of the warm absorber cannot be closer to the nucleus than the NLR. So Phase 2 of the warm absorber must lie somewhere between the NLR and the maximum distance of 1.3 kpc at which the absorber can be kept photoionised. Spatially resolved NLR in nearby Seyferts show diameters of order $10^2$--$10^3$ pc (\citealt{Ost06}). The range of possible distances from the nucleus for the dusty warm absorber (Phase 2) overlaps the range which the NLR is likely to occupy, so the two regions could be co-spatial. On the other hand, the higher-ionised Phase 1, which is not co-spatial with Phase 2 on the grounds of stability analysis and different outflow velocities, has $r< 52$, which places it interior to Phase 2 and the NLR.

\section{Conclusions}
\label{conclusions}
We have presented the first-ever study of the X-ray warm absorber and nuclear obscuration in the Seyfert 1.8 \eso. We conclude that:

\begin{enumerate}

\item The warm absorber which is detected by the \xmm RGS consists of two phases of ionisation, with $\log\xi\approx 3.2$ and 2.3 respectively. The higher-ionised component (Phase 1), with $\NH \approx 1.6 \times 10^{22}$ $\mathrm{cm}^{-2}$, is outflowing with $v \approx -1100\ \kms$; the lower-ionised component (Phase 2), with $\NH \approx 0.5 \times 10^{22}$ $\mathrm{cm}^{-2}$, is outflowing with $v \approx -700\ \kms$.

\item \eso displays a large Balmer decrement (${\rm{\Ha/\Hb} \sim 8}$) relative to the theoretical predictions for the NLR and BLR, implying a significant amount of reddening with ${E(B-V) =}$~0.64\,--\,0.73 and a large column of cold gas, with $\NH =\,$(3.5--5.0)\ $\times 10^{21}\ {\rm{cm}}^{-2}$, in the host galaxy of the AGN. However, the X-ray spectrum is not absorbed by such column of gas; we find an upper limit of $8.9 \times 10^{19}\ {\rm{cm}}^{-2}$ for the neutral hydrogen equivalent column density \NH.

\item Alternative explanations for the discrepancy between the large Balmer decrement and lack of X-ray absorption are discussed. We find partially covering gas clouds or an intrinsically large Balmer decrement are not feasible solutions since the shape of the optical-UV continuum flux (supposed to be thermal emission from the accretion disc) also indicates reddening. We suggest a dusty warm absorber is a viable explanation for the discrepancy, without requiring unusually large dust-to-gas ratio clouds in the ISM of the host galaxy.

\item We show that graphite dust grains can survive the sublimation and sputtering processes in the lower-ionised phase of the warm absorber (Phase 2), and thus can cause the observed optical-UV reddening, whereas the X-rays remain unabsorbed due to lack of neutral column of gas in the ionised absorber. 

\item We have explored the uncertainties in the construction of the SED of \eso due to nuclear obscuration of the source and the effects these have on the results of our warm absorber analysis and the ionisation balance calculations required for photoionisation modelling. The only parameter of the warm absorber outflows which significantly changes as a consequence of using different SEDs is the ionisation parameter. Furthermore, different SEDs result in different thermal stability curves, which determine whether or not the two phases of the warm absorber are in pressure equilibrium. We find that only for the case where the SED is corrected for reddening using the Balmer decrement in the host galaxy and the warm absorber does receive IR radiation from the dusty torus expected to surround the AGN, the two phases of ionisation can co-exist in pressure equilibrium on the thermal stability curve, whereas for other SEDs the two phases are distinct and not in pressure equilibrium. 

\item This work demonstrates the importance of establishing the intrinsic broad-band continuum SED in order to interpret correctly the structure of the warm absorber. We conclude that the two phases of the warm absorber in \eso are most likely separate distinct phases and not in pressure equilibrium; and the dusty phase of the warm absorber is likely to be co-spatial with the NLR. 

\end{enumerate}

\begin{acknowledgements}
This work is based on observations obtained with \xmm, an ESA science mission with instruments and contributions directly funded by ESA member states and the USA (NASA). MM acknowledges the support of a PhD studentship awarded by the UK Science \& Technology Facilities Council (STFC). MM acknowledges useful discussion with K. Wu and C. J. Saxton. We thank the anonymous referee for their useful suggestions and comments.
\end{acknowledgements}

\end{document}